\documentclass[aps,prl, reprint, floatfix,amsmath,amssymb, superscriptaddress,twocolumn]{revtex4-2}

\usepackage[T1]{fontenc}
\usepackage[utf8]{inputenc}
\usepackage{color}
\usepackage{babel}
\usepackage{verbatim}
\usepackage{float}
\usepackage{lmodern}
\usepackage{amsmath,amssymb,dsfont,amstext,amsfonts}
\usepackage{bbold}
\usepackage{mathrsfs}
\usepackage{mathtools}
\usepackage{physics}
\usepackage{hyperref}
\usepackage{lipsum}

\hypersetup{hypertex=true,colorlinks=true,citecolor=blue,urlcolor=blue,linkcolor=blue,anchorcolor=blue}

\renewcommand{\vec}[1]{\mathbf{#1}}

\newcommand{\beginsupplement}{%
    \setcounter{table}{0}
    \renewcommand{\thetable}{S\arabic{table}}%
    \setcounter{figure}{0}
    \renewcommand{\thefigure}{S\arabic{figure}}%
    \setcounter{equation}{0}
    \renewcommand{\theequation}{S\arabic{equation}}%
}

\makeatother

\begin{document}
	\title{
%		Twistronics of rhombohedral graphite
% OVY: This is just a proposal for another title - to be discussed
Electronic states at twist stacking faults in rhombohedral graphite
	}
	% \begin{abstract}
	% 	Topological materials typically feature stable boundary modes that originate from bulk topology.
		
	% \end{abstract}
	% According to a comprehensive inspection of the edge states
	% \begin{comment}
	\author{Xiaoqian Liu}
	\affiliation{Institute of Physics, \'{E}cole Polytechnique F\'{e}d\'{e}rale de Lausanne (EPFL), CH-1015 Lausanne, Switzerland}

    \author{Yifei Guan}
	\affiliation{Institute of Physics, \'{E}cole Polytechnique F\'{e}d\'{e}rale de Lausanne (EPFL), CH-1015 Lausanne, Switzerland}
	
	\author{Oleg V. Yazyev}
	\email{oleg.yazyev@epfl.ch}
	\affiliation{Institute of Physics, \'{E}cole Polytechnique F\'{e}d\'{e}rale de Lausanne (EPFL), CH-1015 Lausanne, Switzerland}

	\date{\today}
 
         \begin{abstract}
            Flat bands in graphitic materials emerged as a platform for realizing tunable correlated physics.
            As a nodal-line semimetal, rhombohedral graphite features flat drumhead surface states in the vicinity of the Dirac points, which carry a nontrivial topological charge.
            We present a comprehensive study on rhombohedral graphite with twist stacking faults.
            Using both the continuum models and the realistic tight-binding models, we show that the twist angle between the graphene layers can tune the interface states at such stacking faults.
            The evolution of interface states originates from the interplay between the moir\'e periodicity and Zak phase topology, predicting the occurrence of nearly flat bands throughout the moir\'e Brillouin zone.
            We further investigate the disorder-induced layer polarization and tunable Chern number for flat band, and characterize the relationship between the disorder strength and Chern number in twisted rhombohedral graphite.
% OVY: what's "band polarization"?
% XQ: It should be "layer polarization".
            % We further characterize the electronic properties of the interface bands through their Berry curvature and valley Chern numbers. 
            % \textcolor{red}{And also discuss the current operator via their quantum geometry.}

        \end{abstract}
	
	\maketitle

% \section{Introduction}
% \lipsum[1-3]

% 1. Twistronics

% 2. Surface states in rhombohedral graphite, also mention the recent works on pentalayer+FCI
% \textcolor{red}{Do not delete my comments.}

% with certain configurations described as 'magic' due to their profound effects. 
% Two-dimensional (2D) materials are notable for their topological properties, because of the incapability of spontaneously breaking a continuous symmetry at finite temperatures \cite{wagner-prl66}, which leads to the stability of the topological phase. These phases rely not on the usual order parameters associated with broken symmetries but on the topological properties of the electronic wave functions \cite{zhang-rmp11}. This gives a particular platform to understand special topological properties, such as quantum spin Hall effect \cite{bernevig-s06, kane-prl05-a}, and topologically protected edge states \cite{topological-insulators-rmp10,fu-prb07}.
Twistronics, an approach unique to 2D materials, dramatically alters the band structure and electronic properties through variations in the twist angle \cite{macdonald-pnas-2011, stephen-twistronic-prb17}. 
% OVY: references to support sentence 1 would not hurt.
% XQ: yes, I added some paper now
Graphene-like systems exhibit exceptional electronic properties due to their Dirac-like charge carriers and linear energy dispersion near the Fermi level~\cite{graphene-rmp}. 
% OVY: If you want to refer to linear dispersion in graphene, these are not correct references.
% XQ: I replaced them, and added the new.
These properties lead to high carrier mobility and suppressed scattering from long-range disorder \cite{das-gra-rmg11}, making graphene a compelling candidate for electronic applications.
% The typical materials, like graphene \cite{kane-prl05-b} and transition metal dichalcogenides \cite{qian-qhe-sci14}, exhibit electrons behaving under topological constraints that lead to robust, fault-tolerant electronic properties resistant to backscattering and defects \cite{topological-insulators-rmp10}. 
Based on the intrinsic properties of graphene, twisted bilayer graphene (TBG) has garnered significant attention because it exhibits a correlated insulating phase \cite{cao-nature18-mott,sharpe-science-19,choi-tbg-stm, marc-tbg-19}, unconventional superconductivity \cite{marc-tbg-19,cao-nature18-supercond, dean-tbg-science19}, and notable topological features \cite{song-tbg-prl19, jpliu-prb19, yang-tbg-prx19}, which sparked widespread interest from researchers.
% in the field of twistronics by introducing a new degree of freedom.
Recently, ferromagnetism, superconductivity, and correlated insulating phases have been observed in more complex twisted heterostructures, such as twisted double bilayer graphene \cite{zhang-tdbg-np20, kim-tdbg-nature20, cao-tdbg-nature20}, twisted trilayer graphene \cite{chen-trilayer-hbn-mott-np19, chen-hbn-trilayer-nature19, chen-trilayer-hbn-qah} as well as rhombohedral multilayer graphene~\cite{shi-rhombo-nature20, han-ferr-nature23, zhou-multi-rhombo-nc24}. 
This indicates that low-energy flat bands that are responsible for the correlated physics in TBG, may also exist in even more complex physical systems eventually extending to the limit of bulk graphite. 
  
% Topological non-trivial flat bands, which are characterized by the odd winding of Wilson loops have already emerged TBG. 
% Consisting with the chirality rhombohedral lattice may be a useful starting point for exploring fractional Chern insulators.
% These flat bands, along with sublattice polarization, contribute to the non-trivial band topology, making the chirality structure a useful starting point for exploring fractional Chern insulators.

% Graphite consists of stacked layers of graphene, typically arranged in a Bernal stacking configuration. But beyond Bernal stacking, the stacking order of multilayer graphene can be rhombohedral \cite{guerrero-trilayer-nanoscale22}.
% Unlike the Bernal stacked structure, the rhombohedral stacked graphene exhibits strong electronic correlation effects like unconventional superconductivity \cite{zhou-sc-nature21,pantaleon-sc-nrp23}, and ferromagnetism \cite{zhou-ferr-rg-nature21,han-ferr-nature23}.
Graphite consists of graphene layers arranged in a Bernal stacking configuration in its thermodynamically stable phase. Beyond Bernal stacking, multilayer graphene can also adopt a metastable rhombohedral stacking arrangement \cite{guerrero-trilayer-nanoscale22}. Unlike Bernal-stacked graphene, multilayer rhombohedral graphene exhibits pronounced electronic correlation effects, including unconventional superconductivity \cite{zhou-sc-nature21,pantaleon-sc-nrp23} and ferromagnetism \cite{zhou-ferr-rg-nature21,han-ferr-nature23}.
% rhombohedral graphene has risen to the forefront of both theoretical and experimental research as an ideal platform for exploring new physics. 
Chirality plays an important role in rhombohedral graphene, leading to several possible states with spontaneously broken symmetry in multi-layer rhombohedral graphene \cite{zan-epc-abc-nc24, shi-rhombo-nature20}.
% so that affects the electronic, optical, magnetic, and topological phenomena. 
Upon increasing the number of layers, rhombohedral graphite is deemed to be protected by chiral symmetry \cite{jiao-zak-prl21,chiu-symmetry-rmp16}, exhibiting a quantized Zak phase and protected surface states \cite{zakphase-prl89, ruiz-rhom-adm23, koshino-nodal-prb20}. 
In the limit of the bulk phase, rhombohedral graphite is a nodal-line semimetal \cite{haering-rg, fang-nodal-prb15, lau-nodel-line-prx21}.
% OVY: I changed the last sentence, but didn't touch the references. Please, cite the Haering paper here and make sure other references correspond to the context.
% XQ: Added.

% the contribution of the flat bands comes from the surface states. 

In this Letter, we investigate the topologically protected flat bands in twisted rhombohedral graphite by using the combination of tight-binding and continuum models. We examine rhombohedral graphite in the chiral limit situation and explore how disorder affects the Chern number of the interface flat band in the rhombohedral systems. 
% OVY: What mean "differen chiral limits"?
% Sorry, I mean it should be ABC-CBA, ABC-ABC in chiral limit situation.
Twisting introduces moir\'e periodicity, which modifies the dispersion of the interface bands through the interplay with the Zak phase topology, and this interface band becomes almost entirely flat when approaching the chiral limit.
% OVY: In this paragraph, you are outlining your work. So, is this your result, or from the reference you are citing?
% XQ: this is my result, I should not cite that paper.
The momentum resolved density of states and the Zak phase pattern at different twist angles in twisted rhombohedral graphite are illustrated.
% The high symmetry point $\Gamma$ is defined as a probe for the LDOS. 
Analysis reveals that the density of states (DOS) in $\Gamma$ point varies with the twist angle, and the flat bands dominate the moir\'e Brillouin Zone (mBZ) near the chiral limit. 
Furthermore, we illustrated the sublattice-polarized flat bands when approaching the chiral limit. 
% remove some "chiral limit"
In the low-energy effective model, the Chern number of flat bands scales linearly with the layer count \cite{jpliu-prx19, han-5graphite-nature23}, but even weak disorder disrupts this scaling since the bandgap exponentially decays with the number of layers. We address disorder effects, showing that the Chern number in twisted rhombohedral graphite eventually converges.

\begin{figure}[htbp]
	\centering
	\includegraphics[width=8.6cm]{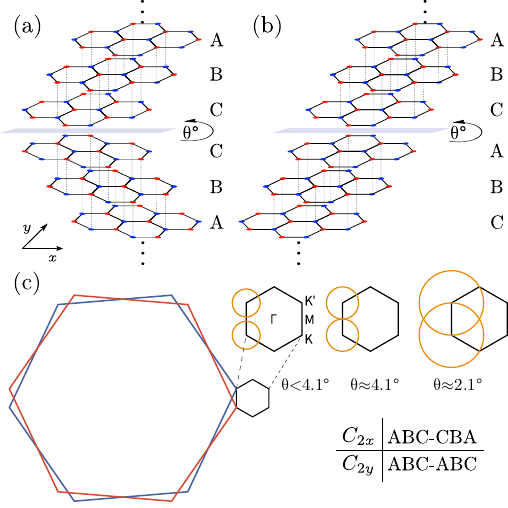}
	\caption{Tiwst stacking faults in rhombohedral graphite with (a) ABC-CBA stacking configuration ($C_{2x}$ symmetry) and (b)  ABC-ABC stacking configuration ($C_{2y}$ symmetry). (c) Moir\'e Brillouin Zone of twisted rhombohedral graphite (black hexagons) at different twist angles. Yellow cicles delimit the region of $\mathcal \mathcal{Z(k_x,k_y)}=\pi$.}\label{fig1}
\end{figure}

% \section{Method}

% Relaxation should be set before calculating the electronic properties and topological properties. The atomic structure of the graphene sheet in ABC and CBA configurations is achieved by the classical force field simulation using LAMMPS\cite{lammps}. The classical force field includes the bond-order potential for covalent bonding as well as the modified Kolmogorov-Crespi potential for interlayer van der Waals interactions. The energy minimization was performed by fire algorithms.

%The Hamiltonian is built with the tight-binding model and the continuum model. 
For the tight-binding (TB) model calculations, we adapted the Slater-Koster formalism \cite{tbmodel-sk-method} for calculating the hopping integrals, which describe intralayer $\pi$ bonds and interlayer $\sigma$ bonds formed by the $p_z$ orbitals of carbon atoms. The atomic positions correspond to the relaxed configurations obtained using LAMMPS \cite{lammps} with the modified Kolmogorov-Crespi potential \cite{lammps-kc-potential} for interlayer van der Waals interactions, and REBO potential \cite{rebo-potential} for intralayer interaction.
% OVY: needs information and the reference on the bond-order potential
% XQ: yes, added.
% The general form of the Hamiltonian, including both contributions, is:

% \begin{equation}
%     \hat{H} = \sum_{i,j}t_{\pi}^{ij}c_i^\dagger c_j+\sum_{i,j}t_{\sigma}^{ij}c_i^\dagger c_j
% \end{equation}

% Explicit expressions for the hopping $t_{\pi}^{ij}$ and $t_{\sigma}^{ij}$ are:

% \begin{equation}
%     \begin{split}
%         t_{\pi}^{ij} = V_{\pi}^0\exp(-\frac{r-a_0}{r_0}) \\
%         t_{\sigma}^{ij} = V_{\sigma}^0\exp(-\frac{r-d_0}{r_0})
%     \end{split}
% \end{equation}

% Following the previous Slater-Koster parametrization, we set $V_{\pi}^0=-2.7 eV$, $V_{\sigma}^0=0.48 eV$, characteristic distances $a=2.46\AA, a_0=1.42\AA$, $d_0=3.35\AA$ and the decay length $r_0=0.184a$. 
% The system is based on the relaxed structure done by LAMMPS \cite{lammps} with the modified Kolmogorov-Crespi potential \cite{lammps-kc-potential} for interlayer van der Waals interactions.

The continuum model of twisted rhombohedral graphene is built starting from the effective model for the multilayer rhombohedral graphene. We focus on the low-energy states near the $K$ point, which we refer to as the $K$ valley, for multilayer rhombohedral graphene, resulting in a $2n\times 2n$ Hamiltonian whose basis is the layer inner product with the sublattice. 
% Expanding in the $\mathbf{K}$ valley, this model will looks like:
% \begin{equation}
%     H_{RG}(\vec{K}+\vec{k}) = \begin{pmatrix}
%         v_F\vec{k}\cdot\sigma & t^\dagger(\vec{k}) & \\
%         t(\vec{k}) & \ddots & t^\dagger(\vec{k}) \\
%          & t(\vec{k}) & v_F\vec{k}\cdot\sigma
%     \end{pmatrix} 
% \end{equation}

% \begin{equation}
%     t(\vec{k}) = \begin{pmatrix}
%         -v_4k & t_1 \\
%         -v_3\bar{k} & -v_4k
%     \end{pmatrix},\quad k,\bar{k} = k_x \pm ik_y
% \end{equation}

% In this Hamiltonian, the terms $v_3$ and $v_4$ are used to represent trigonal warping and particle-hole asymmetry, we neglect these terms to speed up the calculation.
The effect of the twist of the middle graphene layers is described by the twist-induced interlayer coupling $T_\theta(\vec{r})$, which gives

\begin{equation}
    T_\theta(\vec{r}) = \sum T_ne^{-i\vec{q}_n \vec{r}}
\end{equation}
with a moir\'e three-fold star of $\vec{q}_i=2k_D\sin(\theta/2)$, equirotated by $\phi=2\pi/3$

\begin{equation}
    T_n = e^{-i\mathcal{G_\theta^{(\mathrm{n})}}\vec{d}}\hat{\Omega}_\phi^{n-1}
    \begin{pmatrix}
        w_{AA} & w_{AB} \\
        w_{AB} & w_{AA}
    \end{pmatrix}
    \hat{\Omega}_\phi^{1-n}
\end{equation}
with $\hat{\Omega}_\phi=\cos\phi\sigma_x-\sin\phi\sigma_y$. Here, $\mathcal{G}_\theta^{(0)}=0, \mathcal{G}_\theta^{(1)}=\vec{q}_2-\vec{q}_1, \mathcal{G}_\theta^{(2)}=\vec{q}_3-\vec{q}_1$ are the moir\'e reciprocal lattice vectors and $\vec{d}$ is the interlayer displacement\ \cite{macdonald-pnas-2011}.
The twisted rhombohedral graphite (TRG) Hamiltonian thus reads

% \begin{equation}
%     H_{TRG} = \begin{pmatrix}
%         v_F\vec{k}\cdot\sigma & t^\dagger(\vec{k}) &  &  &  \\
%         t(\vec{k}) & \ddots & t^\dagger(\vec{k}) &  &  &  \\
%          & t(\vec{k}) & v_F\vec{k}\cdot\sigma & T_\theta^\dagger(\vec{r})  &  &  \\
%          &  & T_\theta(\vec{r}) & v_F\vec{k}\cdot\sigma & t^\dagger(\vec{k}) \\
%          &  &  & t(\vec{k}) & \ddots & t^\dagger(\vec{k}) \\
%          &  &  &  & t(\vec{k}) & v_F\vec{k}\cdot\sigma
         
%     \end{pmatrix} 
% \end{equation}

\begin{equation}
    H(\vec k) = \begin{pmatrix}
    
        \ddots & t^\dagger(\vec{k}) &  &  \\
        t(\vec{k}) & v_F\vec{k}\cdot\sigma & T_\theta^\dagger(\vec{r})  &  \\
         & T_\theta(\vec{r}) & v_F\vec{k}\cdot\sigma & t^\dagger(\vec{k}) \\
         &  & t(\vec{k}) & \ddots 
         
    \end{pmatrix} 
\end{equation}

More details about the method can be found in~\ref{si_method}.

% \section{Electronic structures of TRG}

Rhombohedral graphite exhibits a distinct stacking sequence that distinguishes it from Bernal-stacked graphite. Unlike Bernal-stacked graphite, where carbon atoms follow the AB stacking sequence, rhombohedral graphite assumes the ABC sequence that breaks inversion symmetry. In turn, this results in the ABC and CBA stacking configurations of different chirality, and one can be transformed into another by a $60^\circ$ rotation. For comparison, when discussing twist stacking faults, we focus on much smaller twist angles of the order of degrees.

Being a three-dimensional nodal-line semimetal, rhombohedral graphite is characterized by the drumhead surface states~\cite{ruiz-rhom-adm23} originating from the nontrivial Zak phase.
Accounting for only the nearest-neighbor coupling between the graphene layers, the system can be modeled as a one-dimensional chain.
The Zak phase $\mathcal{Z}$ is calculated as the integral of the Berry connection over a loop in the mBZ that spans the reciprocal lattice vector, 
% OVY: lattice constant should be present in the integration limits
% XQ: added
\begin{equation}
    \mathcal{Z}(k_x,k_y) = i\int_{k_z=0}^{2\pi/a_z}\left<u(\vec k)|\partial_{k_z}|u(\vec k)\right>dk_z.\label{zak}
    % \mathcal{Z}(k_x,k_y) = i\int_{BZ}\left<u(\vec k)|\partial_{k_z}|u(\vec k)\right>dk_z.\label{zak}
\end{equation}
where $a_z$ is the lattice constant in the $z$-direction, $\left|u(\vec k)\right>$ are the periodic parts of the Bloch functions at momentum $\vec k =(k_x,k_y,k_z)$.
% OVY: "taken over the mBZ" - this is not correct. Please fix.
% we simply remove it, since the function is integrated over kz, people should understand this meaning.
Such topological invariant is quantized in the chiral-symmetric case, i.e. only when the inter-sublattice terms of the Hamiltonian are nonzero.
In such a scenario, $\mathcal Z = 0$ and $\mathcal Z = \pi$ are the $\mathbb Z_2$ topological invariant.

In rhombohedral graphite, the nontrivial $\mathcal{Z}=\pi$ is localized in a narrow region of radius $\delta_k$ around each Dirac cone of the $K$ and $K'$ valley \cite{zakphase-prl89}, with $\delta_k \approx \frac{t_\perp}{v_f \hbar} \approx 0.015 \AA^{-1}$, where $t_\perp$ is the interlayer hopping parameter and $v_f$ is the Fermi velocity.
% When the twist was introduced between the rhombohedral graphite, the change in moment space affected the Zak phase pattern. 
% When the twist was introduced into the rhombohedral graphite, electronic bands near the twisted interface were not only determined by the interplay between moir\'e periodicity but also the Zak phase.
When a twist is introduced in rhombohedral graphite, the electronic states localized at the twist interface are determined by the interplay between moir\'e periodicity and $\delta_k$. One can anticipate a change in the dispersion and topology of these states when $\delta_k$ becomes comparable to the mBZ dimensions $k_D\sin\theta/2$, where $k_D$ is the Dirac-point wave vectors in the continuum model and $\theta$ is the twist angle. 
% The non-trivial Zak phase region increases as the twist angle decreases.
Using the TB model to calculate the momentum-resolved DOS, we demonstrate the interface bands and their relationship to the interplay between $\delta_k$ and the moir\'e periodicity.
% interplay between these two periodicities and resulting interface bands.

Considering the fact that the surface states of rhombohedral graphite are sublattice polarized \cite{han-5graphite-nature23}, the stacking order of the twisted interface determines the coupling regime between the graphite surfaces.
Specifically, the change in the valley momenta is controlled by the stacking order of twisted rhombohedral graphite. 
For the ABC-CBA configuration, as shown in Fig.\ \ref{fig1}(a), the $K$ of the bottom layer matched with the $K'$ in the top layer, and for the ABC-ABC configuration, the $K$ of the bottom layer should match with the $K$ in the top layer. 
Both of the two configurations belong to the $D_3$ symmetry group, but differ by having the $C_{2x}$ and $C_{2y}$ symmetries.
The interface in twisted rhombohedral graphite consists of two separate rhombohedral graphite domains; thus, the chirality and sublattice polarization arising from their stacking sequences must be considered when studying interface properties. Due to the reason that the ABC-CBA configuration provides the flatter interface band when approaching the chiral limit, see~\ref{si_aa-ac}, we use ABC-CBA as the default stacking setup below.
% \textcolor{red}{Compared to the ABC-ABC interface, the ABC-CBA interface implies a uniform alignment across multiple layers that are in the relatively same position. The ABC-CBA configuration increases the connection between the different sublattices.}

% The protection of the Zak phase characterizes the surface state of semi-infinite rhombohedral graphite. 

The interface of the twisted rhombohedral graphite is described by the Green's function 
\begin{equation}
    G(E,\vec{k}) = [E+i\eta-H(\vec{k})-\Sigma_T(\vec{k})-\Sigma_B(\vec{k})]^{-1},\label{green}
\end{equation}
where $\eta$ is a small energy broadening and $H$ is the Hamiltonian of the twisted interface, which is essentially the same as TBG. 
$\Sigma_T$ and $\Sigma_B$ are the self-energy terms from the top and bottom semi-infinite rhombohedral graphite.
We use the momentum resolved density of states $ \rho(E,\vec{k}) = -\frac{1}{\pi}\mathrm{Im}[\mathbf{Tr}G(E,\vec{k})]$ to demonstrate the spectrum of interface localized states.

% Here we use the tight-binding method to construct the momentum-space Green's function.

% \subsection{Evolution of the interface band}

% it is strange that you return to the discussion of zak phase. I would present the sentences below a bit later

\begin{figure}[htbp]
	\centering
	\includegraphics[width=8.6cm]{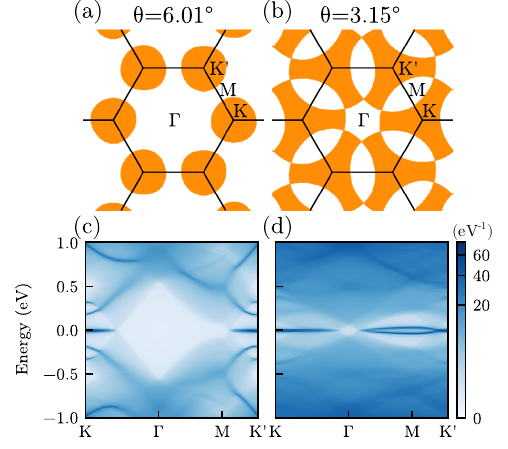}
	\caption{Maps of the Zak phase of rhombohedral graphite for moir\'e periodicities that correspond to twist angle (a) $\theta  = 6.01^{\circ}$ and (b) $\theta  = 3.15^{\circ}$  showing the regions of $\mathcal{Z}=\pi$ in orange. Momentum resolved DOS of twisted rhombohedral graphite at (c) $\theta  = 6.01^{\circ}$ and (d) $\theta  = 3.15^{\circ}$ calculated using the TB model along the $k$-point path $K-\Gamma-M-K'$. }\label{fig2}
\end{figure}

% OVY: "Interplay" is the parasite word of this paper!
% XQ: relation, connection, correspondence? 
% XQ: Would this be good? "To illustrate how the non-trivial Zak-phase region manifests within the mBZ,"?
% To illustrate the interplay between the non-trivial Zak phase region and the mBZ,
To illustrate how the non-trivial Zak-phase region manifests within the mBZ, 
we shall focus on the twist angles around $k_D\sin \theta = \delta_k$ where the $\mathcal Z =\pi $ regions of adjacent moir\'e valleys are expected to overlap. 
% OVY: I do not understand the last sentence. What means this "to be quantized"?
% XQ: "to be clear would be good"?
% OVY: And by the way, which of the two stackings are qwe discussing here?
% XQ: here is not related to the stacking, because the relation between zak phase and mBZ always exist.
To be clear, the $\mathcal{Z}=\pi$ would cancel each other, resulting in the trivial Zak phase $\mathcal{Z}=0$ in the overlapped regions. 
% \com{(what are the angles we use, and why)}
We use two simple cases where the $\mathcal{Z}=\pi$ pattern evolution to illustrate the interplay.
% OVY: I don't understand "pattern interaction".
% XQ: It is the zak phase pattern touch, cancel each other, maybe we can use evolution here?
As shown in Figs.\ \ref{fig2}(a,b), the orange regions represents the $\mathcal{Z}=\pi$, and the blank regions represents the $\mathcal{Z}=0$ at twist angles $6.01^{\circ}$ and $3.15^{\circ}$, respectively.
Figs.\ \ref{fig2}(c,d) show the momentum resolved density of states of the corresponding configurations. 
The bands near the Fermi level in these two configurations are almost flat. 
% OVY: Before DOS was mentioned, now LDOS. Am I right that only LDOS is calculated everywhere? If so, please correct accordingly.
% XQ: DOS refers to the D(E), how many electronic states exist per unit energy, and LDOS refers to the DOS at a certain position. In our case, it is \rho(k,E), so it should be LDOS.
From the Zak phase patterns in Figs.\ \ref{fig2}(a,b), we can easily associate the flat bands at zero energy with the non-trivial Zak phase.
% OVY: I did not understand the text below, so I commented it oput.
% XQ: agreed.
%which means they are zero-mode interface states as well \cite{zurita-zeromode-quantum21}.

% \subsection{Band width versus twist angle}
% The local density of states of the interface results from the Zak phase, and the moir\'e Brillouin zone causes the Zak phase patterns to overlap. 

\begin{figure}[htbp]
	\centering
% OVY: Looking at the figures I would not call this quantity "LDOS". To start with, LDOS can be only positive. Caption was not edited.
% XQ: Yes, see this paper[https://www.science.org/doi/10.1126/sciadv.adi6063], it should be just LDOS.
	\includegraphics[width=8.6cm]{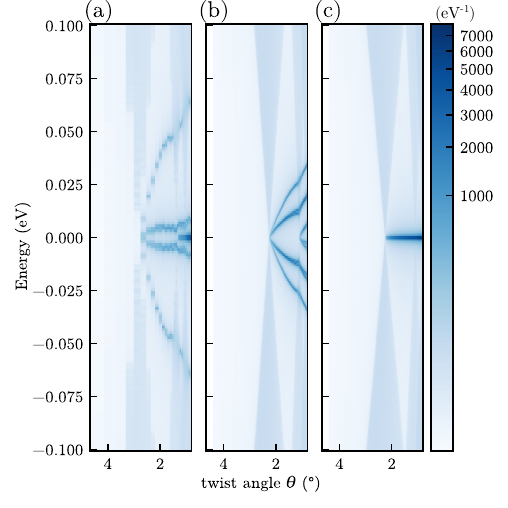}
	\caption{Evolution of the DOS of rhombohedral graphite (semi-infinite) interface states at $\Gamma$ point versus twist angles, the twist angle from $4.41^{\circ}$ to $0.8^{\circ}$; (a) Tight-binding model result; (b) continuum model result, $w_{AA}=20\,$meV; (c) continuum model result, chiral limit, $w_{AA}=0\,$meV.}\label{fig3}
\end{figure}

% OVY: I do not understand the sentence below - did not edit it.
For twist angles around $\theta \approx 2.1^\circ$, as you can see from Fig~\ref{fig3}, only the non-trivial Zak phase ($\mathcal{Z} = \pi$) from the mBZ significantly contributes, simplifying the description of the zero-mode interface states. However, at smaller twist angles, the Zak phase patterns become very complex, respectively resulting in complex dispersion of the interface bands.
In such cases, the trigonal warping orientation that is affected by the stacking order can not be ignored. 
% OVY: I do not understand the sentence below - did not edit it.
To present the situation at the small twist angle and chiral limit, the evolution of the DOS at $\Gamma$ would be adopted.
% The limit of the general flat band depends on the Zak phase overlap. 
% So the chirality of ABC-CBA and ABC-ABC configurations would lead to total different local density of states.

Figure~\ref{fig3} shows the evolution of the DOS of the interface states at the $\Gamma$ point as a function of twist angle. A pronounced change in the DOS occurs around $\theta = 2.6^{\circ}$, corresponding to the angle, at which the Zak phase in momentum space takes the value of $\pi$, reaching the $\Gamma$ point. At this critical angle, the tight-binding calculation reveals interface bands with finite dispersion extending up to 20\ meV. A consistent feature is also observed in the DOS evolution shown in Fig.~\ref{dos}. This energy scale can be contrasted with a typical interaction scale \cite{cao-nature18-mott,xie-tbg_nature19}, where interaction effects are expected to dominate.
% OVY: I agree with Yifei's comment - "discontinuity" is not the right word here. Overall - this statement along with Eq. 6 does not even deserve to be mentioned.
% XQ: I would use discrete to replace discontinuity, would you agree?

% The commensurate angles of TBG are defined by the lattice vectors $(m,n)=mv_1+nv_2$, where the interlayer twist is considered as an operator $\hat{R}(m,n)=(n,m)$, so the twist angle $\theta$ is given by:

% \begin{equation}
%     \theta(n,m) = \arg\left[\frac{m+n/2+\sqrt{3}n/2}{n+m/2+\sqrt{3}m/2}\right]
% \end{equation}

% The flat bands are doubly degenerate. 
Fig~\ref{fig3}(a) shows the tight-binding result. The discrete observed in Fig.\ \ref{fig3}(a) arises from the discrete sampling of twist angles. 
Figs~\ref{fig3}(b) and (c) are the two sublattice continuum model results, which are used to clarify the difference in the chiral limit.
Fig.\ \ref{fig3}(b) shows the finite dispersion with $w_{AA}=20\,$meV, while Fig.\ \ref{fig3}(c) shows the chiral limit result revealing a perfectly flat band at the Fermi level. 
Tight-binding calculation captures interactions beyond nearest neighbors, so the AA coupling would not be zero.
% OVY: Word "crossing" below is ambiguous - please replace with something more clear.
% XQ: Now it is "the overlap between Z=\pi, balabala"
As discussed above, the overlap between $\mathcal{Z}=\pi$ causes the band dispersion, but $\mathcal{Z}=\pi$ itself is the origin of the interface flat bands. So, once the non-trivial Zak phase extends across the mBZ, the corresponding flat band also emerges throughout the entire zone. 
% , the general case of the ABC-ABC configuration. 
% Building on our earlier discussions, the ABC-CBA and the ABC-ABC configurations have different sublattice polarizations in the interface, the ABC-CBA configuration can not reach the chiral limit, and the AA connection can not be totally transferred to the AB connection in the interface, in this sense, the remaining finite AA coupling is introduced in this configuration. 

% The transition from AA to AB is complete when reaching the chiral limit. However, the more realistic situation is the existence of finite AA coupling. In the case of the ABC-CBA configuration, this polarization imposes a tighter constraint on the interface. Consequently, the ABC-CBA scenario is closer to the chiral limit and is more likely to produce flat bands.

% \section{Berry curvature and Chern number}

\begin{figure}[htbp]
	\centering
	\includegraphics[width=8.6cm]{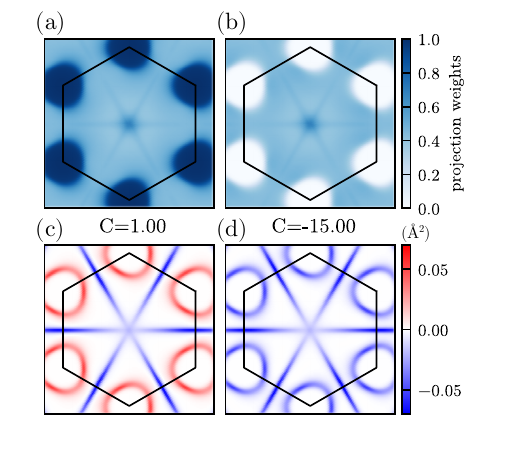}
	\caption{Spectral weight of the (a) interface and (b) surface bands on the interface layers calculated for the 8+8-layer ABC-CBA configuration. (c,d) Berry curvature maps of the respective bands.
}\label{fig_berry}
\end{figure}

To provide a deeper understanding of the electronic properties of the stacking faults in rhombohedral graphite, the Berry phase physics of the interface bands should be taken into account.
While the interface bands are generally not separated from other bands, we retrieve the Berry curvature and valley Chern number by taking the long-sequence limit of twisted rhombohedral multilayer graphene\ \cite{jpliu-prx19,led-chern-prl22}.
We model the twisted rhombohedral graphite by coupling $(M-1)$-layer graphite to the top and bottom layers of TBG. 
The chirally stacked $(M-1)$-layer graphite is described by the degenerate state perturbation Hamiltonian\ \cite{macdonald-chiral-prb08}: 
\begin{equation}
    H_{M-1} = -t_{\perp} \begin{pmatrix}0 & \nu^{M-1}\\
(\nu^{\dagger})^{M-1} & 0
\end{pmatrix},
\end{equation}
where $\nu =\hbar v_f (k_x+i k_y)/t_{\perp}$ represents the effective coupling between the two interface layers of rhombohedral graphene.
In such a setup, there exists a set of flat bands with valley Chern numbers $\pm 2(M-1)$ in the chiral limit, as shown in Fig.~\ref{fig_berry}.
The valley Chern bands are equally localized on the twist interface and the outermost layers of graphite.

The chiral models predict a valley Chern number increasing with $M$, and this dependence is preserved upon continuing the perturbation series\ \cite{slizovskiy-rhombohedral-naturecp19}.
Here, we can ask a question: will the interface states and the surface states remain coherent if $M\rightarrow\infty$? In ideal cases, yes, the coherence length tends to infinity as the temperature approaches absolute zero $\xi(T)\approx\frac{h}{\sqrt{2mk_B T}}$, because there are no thermal disruption to phase coherence. However, in real systems, disorder and quantum fluctuations would limit the coherence length, preventing it from becoming truly infinite. We show below that upon adding non-chiral terms, the interface bands converge to a finite Chern number even in the limit $M\rightarrow\infty$.

% \textcolor{blue}{discuss the possible effect of coherent length on z-direction. eg: use finite-layer setup and add random disorder to the bulk layers, see what happens to the Chern number. This is an analogy of localization transition in one dimension.}

% \textcolor{blue}{add the equation how we define the interface topological numbers[LESS FEASIBLE]
% we need a small and crude definition, not dig deep into that.}

% \textcolor{blue}{[LESS FEASIBLE] quantum geometry: how does the current operator look like at the interface:
% \begin{align}
%     \langle m| \hat j | n \rangle = \langle m | \frac{dH}{dk} |n\rangle
% \end{align}
% non-trivial QG will change the current-current correlation, namely $\sigma$.
% }

The non-chiral terms are just the finite AA-sublattice coupling  $w_{AA}$ between the interface TBG layers. By adding this non-chiral term, we can split the interface state and surface state.
As shown in Fig.\ \ref{fig_berry}(a), a finite $w_{AA}$ significantly modifies the interface-layer spectral weight to the interface band and surface band, causing strong layer polarization. 
As shown in Fig.\ \ref{fig_berry}(a), we focus on the states localized at the interface, finding that such bands always carry $C=\pm 1$ despite the total valley Chern number for interface and surface flat bands $\pm 2(M-1)$.
% which is also present in the perturbative models containing non-chiral terms\ \cite{slizovskiy2019films}.

Moreover, in the chiral limit, all the Chern bands distribute equally between the interface and surfaces. 
Such a delocalization also prevents the Chern number from diverging: the topological number will be constrained by the interlayer coherent length in real systems.

% \textcolor{blue}{Mathematically, the Chern number in the multilayer rhombohedral graphene should scale linearly with the number of layers. In this situation, we expect a couple of coherent states with infinite length, the interface state and surface state. But in reality, the coherent strength should exponentially decay with the distance, so that the interface state and surface state could be decoherent by the random scattering disorder and defects. }

\begin{figure}
    \centering
    \includegraphics[width=8.6cm]{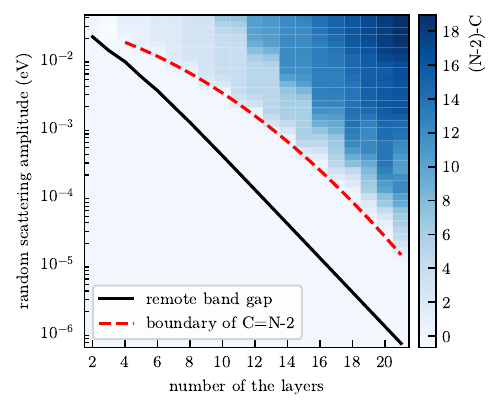}
    \caption{The evolution of total Chern number of the flat bands with the number of layers and the random scattering disorder amplitude. The black line is the remote band gap strength, the color represents the difference between Chern number and its theoretical value $(N-2)-C$.}
    \label{fig_chern}
\end{figure}

To quantitatively investigate these effects, we perform ensemble averaging over independent configurations to analyze how varying disorder strength affects the total Chern number of flat bands. 
We add symmetric intervalley coupling $V=V^\dagger$ as the random scattering between different plane waves to simulate the effect of disorder. 
In the presence of on-site disorder potentials $V_i(r_i)$, the intervalley scattering between moir\'e valleys $\vec{K_{1,2}}$ writes

\begin{equation}
    V_{\chi\bar{\chi}}^{\mathbf{K}_1\mathbf{K}_2} = \sum_{\mathbf{r}}\left<\psi_{\chi,\mathbf{K}_1}(\mathbf{r})\left| V(\mathbf{r})\right| \psi_{\bar{\chi},\mathbf{K}_2}(\mathbf{r})\right>
\end{equation}
where  $\psi=e^{i\mathbf{K}\cdot \vec{r}}$ are the plane-wave basis with $\vec K_i= \vec K_\pm + a_1 \vec G_1 +a_2 \vec G_2$  the indexes of moir\'e valleys,  and $\chi$ identifies the sublattice. $V(\vec{r})=\sum_i V_i \delta(\vec{r_i})$ contains the $\delta$-function scatterings at positions $r_i$. 
Here, we generate the random scattering disorder sequence with a uniform distribution to calculate the statistical average Chern number and evaluate how the total Chern number would change with the disorder. 
% OVY: "distortion" is not a correct word here. Also, please make a normal sentence from the statement below.
% Yes, fixed.
In this model, the layers are arranged in an ABC–CBA configuration.
% XQ: the below sentence is a bit confusing. 
% We refer to the difference between the calculated Chern number C from its theoretical value (N−2).
As shown in Fig.\ \ref{fig_chern}, we refer to the difference between the calculated total Chern number $C$ from its theoretical value $N-2$\ \cite{jpliu-prx19} to make the evolution clear, where $N$ is the number of the layers.
% OVY: Sorry what is C? This was never introduced.
% XQ: We introduced it at line 414.
The random scattering disorder disturbs the linear relation between the number of layers and the total Chern number. The upper limit of 40 meV is critical because the flat band topology for twisted bilayer graphene essentially fails at this disorder strength~\cite{zhangyi-graphene-prb20}.
% OVY: Is the 40 meV limit related to the remote band gap in TBG?
% XQ: Yes, I added a PRB paper here; they mentioned the remote band gap ~ 40 meV.
The weak disorder that is smaller than the remote band gap will not affect the Chern number of the flat band at all. Additionally, the intervalley scattering could mix the Chern number from different bands. This could eventually cause the total Chern number to vanish. By adding the random scattering disorder and taking the statistical average of the total Chern number sequence, we conclude that the Chern number decreases with increasing the disorder amplitude. 

% \section{Conclusion}

In conclusion, we used the Zak phase to explain the emergence of the surface and interface states at twist stacking faults in rhombohedral graphite. The interplay between the moiré Brillouin zone dimensions and the Zak phase offers a physical framework for understanding the formation and evolution of surface and interface states. In the chiral limit, twisted rhombohedral graphite can host degenerate flat bands originating from both surface and interface states. Introducing non-chiral terms, such as the finite AA coupling, we split the degenerate bands and separate the surface and interface states. 

The Chern number of the flat bands of the twisted rhombohedral graphite system is tunable with finite disorder at the surface. As the disorder increases, the Chern number gradually decreases and eventually vanishes, reflecting the breakdown of topological protection. This behavior highlights both the limitations and tunability of flat-band topology, and suggests a promising platform for exploring the interplay between localization, electronic correlations, and topological phases.
% Under specific conditions, such as through doping or applying pressure, these configurations could be a good host of superconductivity, thus guiding the way to unconventional superconducting states.

We thank Jianpeng Liu and Yaroslav Zhumagulov for the fruitful discussions. We acknowledge support by the Swiss National Science Foundation (grant No.204254). Computations were performed at the facilities of the Scientific IT and Application Support Center of EPFL.

\bibliography{liuxq}

@article{han-5graphite-nature23,
  title={Orbital multiferroicity in pentalayer rhombohedral graphene},
  author={Han, Tonghang and Lu, Zhengguang and Scuri, Giovanni and Sung, Jiho and Wang, Jue and Han, Tianyi and Watanabe, Kenji and Taniguchi, Takashi and Fu, Liang and Park, Hongkun and others},
  journal={Nature},
  volume={623},
  number={7985},
  pages={41--47},
  year={2023},
  publisher={Nature Publishing Group UK London}
}

@Article{zakphase-prl89,
  author    = {Zak, J.},
  journal   = {Phys. Rev. Lett.},
  title     = {Berry's phase for energy bands in solids},
  year      = {1989},
  month     = {Jun},
  pages     = {2747--2750},
  volume    = {62},
  doi       = {10.1103/PhysRevLett.62.2747},
  issue     = {23},
  numpages  = {0},
  publisher = {American Physical Society},
}

@article{macdonald-chiral-prb08,
  title = {Chiral decomposition in the electronic structure of graphene multilayers},
  author = {Min, Hongki and MacDonald, A. H.},
  journal = {Phys. Rev. B},
  volume = {77},
  issue = {15},
  pages = {155416},
  numpages = {5},
  year = {2008},
  month = {Apr},
  publisher = {American Physical Society},
  doi = {10.1103/PhysRevB.77.155416},
  url = {https://link.aps.org/doi/10.1103/PhysRevB.77.155416}
}

@Article{chen-trilayer-hbn-mott-np19,
author={Chen, Guorui
and Jiang, Lili
and Wu, Shuang
and Lyu, Bosai
and Li, Hongyuan
and Chittari, Bheema Lingam
and Watanabe, Kenji
and Taniguchi, Takashi
and Shi, Zhiwen
and Jung, Jeil
and Zhang, Yuanbo
and Wang, Feng},
title={Evidence of a gate-tunable Mott insulator in a trilayer graphene moir{\'e} superlattice},
journal={Nat. Phys.},
year={2019},
month={Mar},
day={01},
volume={15},
number={3},
pages={237-241},
doi={10.1038/s41567-018-0387-2}
}

@article{stephen-twistronic-prb17,
  title = {Twistronics: Manipulating the electronic properties of two-dimensional layered structures through their twist angle},
  author = {Carr, Stephen and Massatt, Daniel and Fang, Shiang and Cazeaux, Paul and Luskin, Mitchell and Kaxiras, Efthimios},
  journal = {Phys. Rev. B},
  volume = {95},
  issue = {7},
  pages = {075420},
  numpages = {6},
  year = {2017},
  month = {Feb},
  publisher = {American Physical Society},
  doi = {10.1103/PhysRevB.95.075420},
  url = {https://link.aps.org/doi/10.1103/PhysRevB.95.075420}
}

@article{macdonald-pnas-2011,
  title={Moir{\'e} bands in twisted double-layer graphene},
  author={Bistritzer, Rafi and MacDonald, Allan H},
  journal={Proceedings of the National Academy of Sciences},
  volume={108},
  number={30},
  pages={12233--12237},
  year={2011},
  publisher={National Academy of Sciences}
}

@Article{slizovskiy-rhombohedral-naturecp19,
  author    = {Slizovskiy, Sergey and McCann, Edward and Koshino, Mikito and Fal’ko, Vladimir I},
  journal   = {Communications physics},
  title     = {Films of rhombohedral graphite as two-dimensional topological semimetals},
  year      = {2019},
  number    = {1},
  pages     = {164},
  volume    = {2},
  publisher = {Nature Publishing Group UK London},
}

@article{ruiz-rhom-adm23,
author = {Garcia-Ruiz, Aitor and Slizovskiy, Sergey and Fal'ko, Vladimir I.},
title = {Flat Bands for Electrons in Rhombohedral Graphene Multilayers with a Twin Boundary},
journal = {Advanced Materials Interfaces},
volume = {10},
number = {7},
pages = {2202221},
doi = {https://doi.org/10.1002/admi.202202221},
year = {2023}
}

@article{xie-tbg_nature19,
  title={Spectroscopic signatures of many-body correlations in magic-angle twisted bilayer graphene},
  author={Xie, Yonglong and Lian, Biao and J{\"a}ck, Berthold and Liu, Xiaomeng and Chiu, Cheng-Li and Watanabe, Kenji and Taniguchi, Takashi and Bernevig, B Andrei and Yazdani, Ali},
  journal={Nature},
  volume={572},
  number={7767},
  pages={101--105},
  year={2019},
  publisher={Nature Publishing Group UK London}
}

@article{zhou-multi-rhombo-nc24,
  title={Layer-polarized ferromagnetism in rhombohedral multilayer graphene},
  author={Zhou, Wenqiang and Ding, Jing and Hua, Jiannan and Zhang, Le and Watanabe, Kenji and Taniguchi, Takashi and Zhu, Wei and Xu, Shuigang},
  journal={Nature Communications},
  volume={15},
  number={1},
  pages={2597},
  year={2024},
  publisher={Nature Publishing Group UK London}
}

@article{cao-nature18-mott,
  title={Correlated insulator behaviour at half-filling in magic-angle graphene superlattices},
  author={Cao, Yuan and Fatemi, Valla and Demir, Ahmet and Fang, Shiang and Tomarken, Spencer L and Luo, Jason Y and Sanchez-Yamagishi, Javier D and Watanabe, Kenji and Taniguchi, Takashi and Kaxiras, Efthimios and others},
  journal={Nature},
  volume={556},
  number={7699},
  pages={80},
  year={2018},
  publisher={Nature Publishing Group}
}

@article{cao-nature18-supercond,
  title={Unconventional superconductivity in magic-angle graphene superlattices},
  author={Cao, Yuan and Fatemi, Valla and Fang, Shiang and Watanabe, Kenji and Taniguchi, Takashi and Kaxiras, Efthimios and Jarillo-Herrero, Pablo},
  journal={Nature},
  volume={556},
  number={7699},
  pages={43},
  year={2018},
  publisher={Nature Publishing Group}
}

@article{graphene-rmp,
  title = {The electronic properties of graphene},
  author = {Castro Neto, A. H. and Guinea, F. and Peres, N. M. R. and Novoselov, K. S. and Geim, A. K.},
  journal = {Rev. Mod. Phys.},
  volume = {81},
  issue = {1},
  pages = {109--162},
  numpages = {0},
  year = {2009},
  month = {Jan},
  publisher = {American Physical Society},
  doi = {10.1103/RevModPhys.81.109},
}

@article{rebo-potential,
  title={A second-generation reactive empirical bondorder (REBO) potential energy expression for hydrocarbons},
  author={Brenner, Donald W and Shenderova, Olga A and Harrison, Judith A and Stuart, Steven J and Ni, Boris and Sinnott, Susan B},
  journal={Journal of Physics: Condensed Matter},
  volume={14},
  number={4},
  pages={783},
  year={2002},
  publisher={IOP Publishing}
}

@article{tb-graphene,
  title = {Tight-binding description of graphene},
  author = {Reich, S. and Maultzsch, J. and Thomsen, C. and Ordej\'on, P.},
  journal = {Phys. Rev. B},
  volume = {66},
  issue = {3},
  pages = {035412},
  numpages = {5},
  year = {2002},
  month = {Jul},
  publisher = {American Physical Society},
  doi = {10.1103/PhysRevB.66.035412},
  url = {https://link.aps.org/doi/10.1103/PhysRevB.66.035412}
}

@article{zhangyi-graphene-prb20,
  title = {Correlated insulating phases of twisted bilayer graphene at commensurate filling fractions: A Hartree-Fock study},
  author = {Zhang, Yi and Jiang, Kun and Wang, Ziqiang and Zhang, Fuchun},
  journal = {Phys. Rev. B},
  volume = {102},
  issue = {3},
  pages = {035136},
  numpages = {9},
  year = {2020},
  month = {Jul},
  publisher = {American Physical Society},
  doi = {10.1103/PhysRevB.102.035136},
  url = {https://link.aps.org/doi/10.1103/PhysRevB.102.035136}
}

@ARTICLE{sharpe-science-19,
       author = {{Sharpe}, Aaron L. and {Fox}, Eli J. and {Barnard}, Arthur W. and {Finney}, Joe and {Watanabe}, Kenji and {Taniguchi}, Takashi and
{Kastner}, M.~A. and {Goldhaber-Gordon}, David},
       title = "{Emergent ferromagnetism near three-quarters filling in twisted bilayer graphene}",
      journal = {Science},
         year = "2019",
        month = "Aug",
       volume = {365},
       number = {6453},
        pages = {605-608},
          doi = {10.1126/science.aaw3780},
}

@article{choi-tbg-stm,
  title={Imaging Electronic Correlations in Twisted Bilayer Graphene near the Magic Angle},
  author={Choi, Youngjoon and Kemmer, Jeannette and Peng, Yang and Thomson, Alex and Arora, Harpreet and Polski, Robert and Zhang, Yiran and Ren, Hechen and Alicea, Jason and Refael, Gil and others},
  journal={arXiv preprint arXiv:1901.02997},
  year={2019}
}

@article{marc-tbg-19,
author = {Codecido, Emilio and Wang, Qiyue and Koester, Ryan and Che, Shi and Tian, Haidong and Lv, Rui and Tran, Son and Watanabe, Kenji and Taniguchi, Takashi and Zhang, Fan and Bockrath, Marc and Lau, Chun Ning},
title = {Correlated insulating and superconducting states in twisted bilayer graphene below the magic angle},
volume = {5}, 
number = {9},
year = {2019}, 
doi = {10.1126/sciadv.aaw9770}, 
publisher = {American Association for the Advancement of Science}, 
journal = {Science Advances}
}

@article{jpliu-prb19,
  title = {Pseudo Landau level representation of twisted bilayer graphene: Band topology and implications on the correlated insulating phase},
  author = {Liu, Jianpeng and Liu, Junwei and Dai, Xi},
  journal = {Phys. Rev. B},
  volume = {99},
  issue = {15},
  pages = {155415},
  numpages = {9},
  year = {2019},
  month = {Apr},
  publisher = {American Physical Society},
  doi = {10.1103/PhysRevB.99.155415},
}

@article{jpliu-prx19,
  title = {Quantum Valley Hall Effect, Orbital Magnetism, and Anomalous Hall Effect in Twisted Multilayer Graphene Systems},
  author = {Liu, Jianpeng and Ma, Zhen and Gao, Jinhua and Dai, Xi},
  journal = {Phys. Rev. X},
  volume = {9},
  issue = {3},
  pages = {031021},
  numpages = {14},
  year = {2019},
  month = {Aug},
  publisher = {American Physical Society},
  doi = {10.1103/PhysRevX.9.031021},
}

@article{tbmodel-sk-method,
  title = {Simplified LCAO Method for the Periodic Potential Problem},
  author = {Slater, J. C. and Koster, G. F.},
  journal = {Phys. Rev.},
  volume = {94},
  issue = {6},
  pages = {1498--1524},
  numpages = {0},
  year = {1954},
  month = {Jun},
  publisher = {American Physical Society},
  doi = {10.1103/PhysRev.94.1498},
  url = {https://link.aps.org/doi/10.1103/PhysRev.94.1498}
}

@article{lammps-kc-potential,
  title = {Registry-dependent interlayer potential for graphitic systems},
  author = {Kolmogorov, Aleksey N. and Crespi, Vincent H.},
  journal = {Phys. Rev. B},
  volume = {71},
  issue = {23},
  pages = {235415},
  numpages = {6},
  year = {2005},
  month = {Jun},
  publisher = {American Physical Society},
  doi = {10.1103/PhysRevB.71.235415},
  url = {https://link.aps.org/doi/10.1103/PhysRevB.71.235415}
}

@article{dean-tbg-science19,
  title={Tuning superconductivity in twisted bilayer graphene},
  author={Yankowitz, Matthew and Chen, Shaowen and Polshyn, Hryhoriy and Zhang, Yuxuan and Watanabe, K and Taniguchi, T and Graf, David and Young, Andrea F and Dean, Cory R},
  journal={Science},
  volume={363},
  number={6431},
  pages={1059--1064},
  year={2019},
  publisher={American Association for the Advancement of Science}
}

@article{song-tbg-prl19,
  title = {All Magic Angles in Twisted Bilayer Graphene are Topological},
  author = {Song, Zhida and Wang, Zhijun and Shi, Wujun and Li, Gang and Fang, Chen and Bernevig, B. Andrei},
  journal = {Phys. Rev. Lett.},
  volume = {123},
  issue = {3},
  pages = {036401},
  numpages = {6},
  year = {2019},
  month = {Jul},
  publisher = {American Physical Society},
  doi = {10.1103/PhysRevLett.123.036401},
}

@article{yang-tbg-prx19,
  title = {Failure of Nielsen-Ninomiya Theorem and Fragile Topology in Two-Dimensional Systems with Space-Time Inversion Symmetry: Application to Twisted Bilayer Graphene at Magic Angle},
  author = {Ahn, Junyeong and Park, Sungjoon and Yang, Bohm-Jung},
  journal = {Phys. Rev. X},
  volume = {9},
  issue = {2},
  pages = {021013},
  numpages = {26},
  year = {2019},
  month = {Apr},
  publisher = {American Physical Society},
  doi = {10.1103/PhysRevX.9.021013},
}

@Article{chen-hbn-trilayer-nature19,
author={Chen, Guorui
and Sharpe, Aaron L.
and Gallagher, Patrick
and Rosen, Ilan T.
and Fox, Eli J.
and Jiang, Lili
and Lyu, Bosai
and Li, Hongyuan
and Watanabe, Kenji
and Taniguchi, Takashi
and Jung, Jeil
and Shi, Zhiwen
and Goldhaber-Gordon, David
and Zhang, Yuanbo
and Wang, Feng},
title={Signatures of tunable superconductivity in a trilayer graphene moir{\'e} superlattice},
journal={Nature},
year={2019},
volume={572},
number={7768},
pages={215-219},
issn={1476-4687},
doi={10.1038/s41586-019-1393-y},
}

@Article{cao-tdbg-nature20,
author={Cao, Yuan
and Rodan-Legrain, Daniel
and Rubies-Bigorda, Oriol
and Park, Jeong Min
and Watanabe, Kenji
and Taniguchi, Takashi
and Jarillo-Herrero, Pablo},
title={Tunable correlated states and spin-polarized phases in twisted bilayer--bilayer graphene},
journal={Nature},
year={2020},
month={May},
day={06},
issn={1476-4687},
doi={10.1038/s41586-020-2260-6},
}

@Article{zhang-tdbg-np20,
author={Shen, Cheng
and Chu, Yanbang
and Wu, QuanSheng
and Li, Na
and Wang, Shuopei
and Zhao, Yanchong
and Tang, Jian
and Liu, Jieying
and Tian, Jinpeng
and Watanabe, Kenji
and Taniguchi, Takashi
and Yang, Rong
and Meng, Zi Yang
and Shi, Dongxia
and Yazyev, Oleg V.
and Zhang, Guangyu},
title={Correlated states in twisted double bilayer graphene},
journal={Nature Physics},
year={2020},
month={May},
day={01},
volume={16},
number={5},
pages={520-525},
issn={1745-2481},
doi={10.1038/s41567-020-0825-9},
}

@article{chen-trilayer-hbn-qah,
  title={Tunable correlated chern insulator and ferromagnetism in a moir{\'e} superlattice},
  author={Chen, Guorui and Sharpe, Aaron L and Fox, Eli J and Zhang, Ya-Hui and Wang, Shaoxin and Jiang, Lili and Lyu, Bosai and Li, Hongyuan and Watanabe, Kenji and Taniguchi, Takashi and others},
  journal={Nature},
  volume={579},
  number={7797},
  pages={56--61},
  year={2020},
  publisher={Nature Publishing Group}
}

@Article{kim-tdbg-nature20,
author={Liu, Xiaomeng
and Hao, Zeyu
and Khalaf, Eslam
and Lee, Jong Yeon
and Ronen, Yuval
and Yoo, Hyobin
and Haei Najafabadi, Danial
and Watanabe, Kenji
and Taniguchi, Takashi
and Vishwanath, Ashvin
and Kim, Philip},
title={Tunable spin-polarized correlated states in twisted double bilayer graphene},
journal={Nature},
year={2020},
month={Jul},
day={01},
volume={583},
number={7815},
pages={221-225},
doi={10.1038/s41586-020-2458-7}
}

@Article{lammps,
	author = "A. P. Thompson and H. M. Aktulga and R. Berger and 
	D. S. Bolintineanu and W. M. Brown and P. S. Crozier and
	P. J. in 't Veld and A. Kohlmeyer and S. G. Moore and T. D. Nguyen and
	R. Shan and M. J. Stevens and J. Tranchida and C. Trott and S. J. Plimpton",
	title = "{LAMMPS} - a flexible simulation tool for
	particle-based materials modeling at the 
	atomic, meso, and continuum scales",
	journal = "Comp. Phys. Comm.",
	volume =  "271",
	pages =   "108171",
	year =    "2022",
	doi = "10.1016/j.cpc.2021.108171"
}

@article{haering-rg,
  title={Band structure of rhombohedral graphite},
  author={Haering, RR},
  journal={Canadian Journal of Physics},
  volume={36},
  number={3},
  pages={352--362},
  year={1958},
  publisher={NRC Research Press Ottawa, Canada}
}

@article{das-gra-rmg11,
  title = {Electronic transport in two-dimensional graphene},
  author = {Das Sarma, S. and Adam, Shaffique and Hwang, E. H. and Rossi, Enrico},
  journal = {Rev. Mod. Phys.},
  volume = {83},
  issue = {2},
  pages = {407--470},
  numpages = {0},
  year = {2011},
  month = {May},
  publisher = {American Physical Society},
  doi = {10.1103/RevModPhys.83.407},
  url = {https://link.aps.org/doi/10.1103/RevModPhys.83.407}
}

@article{quanshen-rhombohedral-prb24,
  title = {Moir\'e fractional Chern insulators. II. First-principles calculations and continuum models of rhombohedral graphene superlattices},
  author = {Herzog-Arbeitman, Jonah and Wang, Yuzhi and Liu, Jiaxuan and Tam, Pok Man and Qi, Ziyue and Jia, Yujin and Efetov, Dmitri K. and Vafek, Oskar and Regnault, Nicolas and Weng, Hongming and Wu, Quansheng and Bernevig, B. Andrei and Yu, Jiabin},
  journal = {Phys. Rev. B},
  volume = {109},
  issue = {20},
  pages = {205122},
  numpages = {46},
  year = {2024},
  month = {May},
  publisher = {American Physical Society},
  doi = {10.1103/PhysRevB.109.205122},
  url = {https://link.aps.org/doi/10.1103/PhysRevB.109.205122}
}

@Article{zan-epc-abc-nc24,
  author    = {Zan, Xiaozhou and Guo, Xiangdong and Deng, Aolin and Huang, Zhiheng and Liu, Le and Wu, Fanfan and Yuan, Yalong and Zhao, Jiaojiao and Peng, Yalin and Li, Lu and Zhang, Yangkun and Li, Xiuzhen and Zhu, Jundong and Dong, Jingwei and Shi, Dongxia and Yang, Wei and Yang, Xiaoxia and Shi, Zhiwen and Du, Luojun and Dai, Qing and Zhang, Guangyu},
  journal   = {Nature Communications},
  title     = {Electron/infrared-phonon coupling in ABC trilayer graphene},
  year      = {2024},
  issn      = {2041-1723},
  month     = feb,
  number    = {1},
  volume    = {15},
  doi       = {10.1038/s41467-024-46129-7},
  publisher = {Springer Science and Business Media LLC},
}

@Article{zhou-ferr-rg-nature21,
  author    = {Zhou, Haoxin and Xie, Tian and Ghazaryan, Areg and Holder, Tobias and Ehrets, James R. and Spanton, Eric M. and Taniguchi, Takashi and Watanabe, Kenji and Berg, Erez and Serbyn, Maksym and Young, Andrea F.},
  journal   = {Nature},
  title     = {Half- and quarter-metals in rhombohedral trilayer graphene},
  year      = {2021},
  issn      = {1476-4687},
  month     = sep,
  number    = {7881},
  pages     = {429--433},
  volume    = {598},
  doi       = {10.1038/s41586-021-03938-w},
  publisher = {Springer Science and Business Media LLC},
}

@Article{zhou-sc-nature21,
  author    = {Zhou, Haoxin and Xie, Tian and Taniguchi, Takashi and Watanabe, Kenji and Young, Andrea F.},
  journal   = {Nature},
  title     = {Superconductivity in rhombohedral trilayer graphene},
  year      = {2021},
  issn      = {1476-4687},
  month     = sep,
  number    = {7881},
  pages     = {434--438},
  volume    = {598},
  doi       = {10.1038/s41586-021-03926-0},
  publisher = {Springer Science and Business Media LLC},
}

@Article{pantaleon-sc-nrp23,
  author    = {Pantaleón, Pierre A. and Jimeno-Pozo, Alejandro and Sainz-Cruz, Héctor and Phong, Võ Tiến and Cea, Tommaso and Guinea, Francisco},
  journal   = {Nature Reviews Physics},
  title     = {Superconductivity and correlated phases in non-twisted bilayer and trilayer graphene},
  year      = {2023},
  issn      = {2522-5820},
  month     = apr,
  number    = {5},
  pages     = {304--315},
  volume    = {5},
  doi       = {10.1038/s42254-023-00575-2},
  publisher = {Springer Science and Business Media LLC},
}

@Article{shi-rhombo-nature20,
  author    = {Shi, Yanmeng and Xu, Shuigang and Yang, Yaping and Slizovskiy, Sergey and Morozov, Sergey V. and Son, Seok-Kyun and Ozdemir, Servet and Mullan, Ciaran and Barrier, Julien and Yin, Jun and Berdyugin, Alexey I. and Piot, Benjamin A. and Taniguchi, Takashi and Watanabe, Kenji and Fal’ko, Vladimir I. and Novoselov, Kostya S. and Geim, A. K. and Mishchenko, Artem},
  journal   = {Nature},
  title     = {Electronic phase separation in multilayer rhombohedral graphite},
  year      = {2020},
  issn      = {1476-4687},
  month     = aug,
  number    = {7820},
  pages     = {210--214},
  volume    = {584},
  doi       = {10.1038/s41586-020-2568-2},
  publisher = {Springer Science and Business Media LLC},
}

@article{lau-nodel-line-prx21,
  title = {Designing Three-Dimensional Flat Bands in Nodal-Line Semimetals},
  author = {Lau, Alexander and Hyart, Timo and Autieri, Carmine and Chen, Anffany and Pikulin, Dmitry I.},
  journal = {Phys. Rev. X},
  volume = {11},
  issue = {3},
  pages = {031017},
  numpages = {10},
  year = {2021},
  month = {Jul},
  publisher = {American Physical Society},
  doi = {10.1103/PhysRevX.11.031017},
  url = {https://link.aps.org/doi/10.1103/PhysRevX.11.031017}
}

@article{chiu-symmetry-rmp16,
  title={Classification of topological quantum matter with symmetries},
  author={Chiu, Ching-Kai and Teo, Jeffrey CY and Schnyder, Andreas P and Ryu, Shinsei},
  journal={Reviews of Modern Physics},
  volume={88},
  number={3},
  pages={035005},
  year={2016},
  publisher={APS}
}

@article{jiao-zak-prl21,
  title={Experimentally detecting quantized Zak phases without chiral symmetry in photonic lattices},
  author={Jiao, Zhi-Qiang and Longhi, Stefano and Wang, Xiao-Wei and Gao, Jun and Zhou, Wen-Hao and Wang, Yao and Fu, Yu-Xuan and Wang, Li and Ren, Ruo-Jing and Qiao, Lu-Feng and others},
  journal={Physical Review Letters},
  volume={127},
  number={14},
  pages={147401},
  year={2021},
  publisher={APS}
}

@article{guerrero-trilayer-nanoscale22,
  title={Rhombohedral trilayer graphene is more stable than its Bernal counterpart},
  author={Guerrero-Avil{\'e}s, Ra{\'u}l and Pelc, M and Geisenhof, Fabian R and Weitz, R Thomas and Ayuela, Andr{\'e}s},
  journal={Nanoscale},
  volume={14},
  number={43},
  pages={16295--16302},
  year={2022},
  publisher={Royal Society of Chemistry}
}

@article{koshino-nodal-prb20,
  title = {Electronic properties of a graphyne-$N$ monolayer and its multilayer: Even-odd effect and topological nodal line semimetalic phases},
  author = {Kawakami, Takuto and Nomura, Takafumi and Koshino, Mikito},
  journal = {Phys. Rev. B},
  volume = {102},
  issue = {11},
  pages = {115421},
  numpages = {15},
  year = {2020},
  month = {Sep},
  publisher = {American Physical Society},
  doi = {10.1103/PhysRevB.102.115421},
  url = {https://link.aps.org/doi/10.1103/PhysRevB.102.115421}
}

@article{fang-nodal-prb15,
  title = {Topological nodal line semimetals with and without spin-orbital coupling},
  author = {Fang, Chen and Chen, Yige and Kee, Hae-Young and Fu, Liang},
  journal = {Phys. Rev. B},
  volume = {92},
  issue = {8},
  pages = {081201},
  numpages = {5},
  year = {2015},
  month = {Aug},
  publisher = {American Physical Society},
  doi = {10.1103/PhysRevB.92.081201},
  url = {https://link.aps.org/doi/10.1103/PhysRevB.92.081201}
}

@Article{han-ferr-nature23,
  author    = {Han, Tonghang and Lu, Zhengguang and Scuri, Giovanni and Sung, Jiho and Wang, Jue and Han, Tianyi and Watanabe, Kenji and Taniguchi, Takashi and Fu, Liang and Park, Hongkun and Ju, Long},
  journal   = {Nature},
  title     = {Orbital multiferroicity in pentalayer rhombohedral graphene},
  year      = {2023},
  issn      = {1476-4687},
  month     = oct,
  number    = {7985},
  pages     = {41--47},
  volume    = {623},
  date      = {2023-10-18},
  day       = {18},
  doi       = {10.1038/s41586-023-06572-w},
  publisher = {Springer Science and Business Media LLC},
}

@Article{led-chern-prl22,
  author    = {Ledwith, Patrick J and Vishwanath, Ashvin and Khalaf, Eslam},
  journal   = {Physical Review Letters},
  title     = {Family of ideal chern flatbands with arbitrary chern number in chiral twisted graphene multilayers},
  year      = {2022},
  number    = {17},
  pages     = {176404},
  volume    = {128},
  publisher = {APS},
}

\widetext
\clearpage

\begin{center}
\textbf{\large Supplementary Information of ``Electronic states at twist stacking faults in rhombohedral graphite''}
\end{center}

\beginsupplement

\pagenumbering{roman}

% OVY: For some reason, "References" appear as the fist item in the egenrated table of contents. This needs to be corrected, but in any case, the supporting information should be submitted as a separate file.

\tableofcontents

\vspace{12pt}

\section{Methods}\label{si_method}

The Hamiltonian of the tight-binding model is

\begin{equation}
    \hat{H} = \sum_{i,j}t_{\pi}^{ij}c_i^\dagger c_j+\sum_{i,j}t_{\sigma}^{ij}c_i^\dagger c_j .
\end{equation}

Explicit expressions for the hopping parameters $t_{\pi}^{ij}$ and $t_{\sigma}^{ij}$ are

\begin{equation}
    \begin{split}
        t_{\pi}^{ij} = V_{\pi}^0\exp(-\frac{r-a_0}{r_0}) , \\
        t_{\sigma}^{ij} = V_{\sigma}^0\exp(-\frac{r-d_0}{r_0}) .
    \end{split}
\end{equation}

Following the previous Slater-Koster parametrization, 
% OVY: If you say previous - give a reference!
% XQ: Yes, I cite the paper here.
we set $V_{\pi}^0=-2.7$~eV, $V_{\sigma}^0=0.48$~eV \cite{tb-graphene}, characteristic distances $a=2.46 \AA, a_0=1.42 \AA$, $d_0=3.35 \AA$ and the decay length $r_0=0.184a$. 

The continuum model of twisted rhombohedral graphene is built starting from the effective model for the multi-layer rhombohedral graphene. We focus on the low-energy states near the $K$ point, which we refer to as the $K$ valley, for multi-layer rhombohedral graphene, resulting in a $2n\times 2n$ Hamiltonian whose basis is the layer inner product with the sublattice \cite{quanshen-rhombohedral-prb24}. 
Expanding in the $K$ valley, one obtains

\begin{equation}
    H_{RG}(\vec{K}+\vec{k}) = \begin{pmatrix}
        v_F\vec{k}\cdot\sigma & t^\dagger(\vec{k}) & \\
        t(\vec{k}) & \ddots & t^\dagger(\vec{k}) \\
         & t(\vec{k}) & v_F\vec{k}\cdot\sigma
    \end{pmatrix} ,
\end{equation}

\begin{equation}
    t(\vec{k}) = \begin{pmatrix}
        -v_4k & t_1 \\
        -v_3\bar{k} & -v_4k
    \end{pmatrix},\quad k,\bar{k} = k_x \pm ik_y .
\end{equation}

In this Hamiltonian, the terms $v_3$ and $v_4$ used to represent trigonal warping and particle-hole asymmetry are neglected.
The effect of the twist is described by the coupling $T_\theta(\vec{r})$, which gives

\begin{equation}
    T_\theta(\vec{r}) = \sum T_ne^{-i\vec{q}_n \vec{r}}
\end{equation}
with a moir\'e three-fold star $\vec{q}_i=2k_D\sin(\theta/2)$ equirotated by $\phi=2\pi/3$ and

\begin{equation}
    T_n = e^{-i\mathcal{G_\theta^{(\mathrm{n})}}\vec{d}}\hat{\Omega}_\phi^{n-1}
    \begin{pmatrix}
        w_{AA} & w_{AB} \\
        w_{AB} & w_{AA}
    \end{pmatrix}
    \hat{\Omega}_\phi^{1-n}
\end{equation}
with $\hat{\Omega}_\phi=\cos\phi\sigma_x-\sin\phi\sigma_y$. Here, $\mathcal{G}_\theta^{(0)}=0, \mathcal{G}_\theta^{(1)}=\vec{q}_2-\vec{q}_1, \mathcal{G}_\theta^{(2)}=\vec{q}_3-\vec{q}_1$ are the moir\'e reciprocal lattice vectors and $\vec{d}$ is the interlayer displacement\ \cite{macdonald-pnas-2011}.
The TRG Hamiltonian thus reads

\begin{equation}
    H_{TRG} = \begin{pmatrix}
        v_F\vec{k}\cdot\sigma & t^\dagger(\vec{k}) &  &  &  \\
        t(\vec{k}) & \ddots & t^\dagger(\vec{k}) &  &  &  \\
         & t(\vec{k}) & v_F\vec{k}\cdot\sigma & T_\theta^\dagger(\vec{r})  &  &  \\
         &  & T_\theta(\vec{r}) & v_F\vec{k}\cdot\sigma & t^\dagger(\vec{k}) \\
         &  &  & t(\vec{k}) & \ddots & t^\dagger(\vec{k}) \\
         &  &  &  & t(\vec{k}) & v_F\vec{k}\cdot\sigma
         
    \end{pmatrix} 
\end{equation}

Another way to build the twisted rhombohedral graphene is via

\begin{equation}
    H_{TRG} = 
    \begin{pmatrix}
    H_{TBG} & H_{int} \\
    H_{int} & H_{NLG} 
    \end{pmatrix} .
\end{equation}

Here, the TBG block is constructed in the same way as described above, and the $N$-layer graphene block is based on the $n$-th order perturbation theory \cite{macdonald-chiral-prb08}

\begin{equation}
    H_{NLG}^n=-t_\perp
    \begin{pmatrix}
        0 & (v_F/t_\perp\vec{k}_Me^{i(\phi-\theta)})^n \\
        (v_F/t_\perp\vec{k}_Me^{-i(\phi-\theta)})^n & 0
    \end{pmatrix}+\eta\sigma_z ,
\end{equation}

where $\vec{k}_M=\sqrt{k_x^2+k_y^2},\phi=\arg(k_x+ik_y)$, $\theta$ is the twist angle, $\eta$ is a small finite mass. It is necessary to mention, this model highly relies on $t_\perp$, so it only validates around the chiral limit.

\section{Band structure and bandwidth evolution from continuum model}\label{si_band}

\begin{figure}[htbp]
    \centering
    \includegraphics[width=14.6cm]{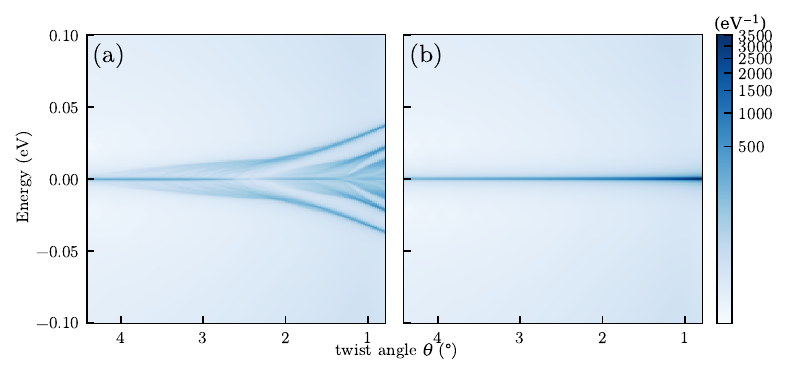}
    \caption{The density of state (DOS) of twisted rhombohedral graphite as a function of $\theta$. (a) and (b), $w_{AA}= 20$ meV and $0$ meV, respectively.}
    \label{dos}
\end{figure}

As shown in Fig.~\ref{dos}(a), the signature of the critical twist angle at $\theta \approx 2.6^\circ$ is not sharply resolved in the total DOS. At this angle, the Zak phase in momentum space takes the value of $\pi$ begins to approach the Brillouin zone center, and the flat bands at the $\Gamma$ point gradually become dispersive. Notably, before reaching this critical angle, the Zak phase in momentum space takes the value of $\pi$ has already started to overlap at the $M$ point. At the critical angle, the bandwidth is approximately 20 meV. This energy scale is well-suited for exploring strongly correlated topological phases, where interaction effects are expected to dominate.
More generally, when finite AA coupling is included, electrons tend to delocalize, whereas in the chiral-limit case, see Fig.~\ref{dos}(b), the bands collapse into highly degenerate, nearly perfectly flat bands.

\begin{figure}[htbp]
    \centering
    \includegraphics[width=14.6cm]{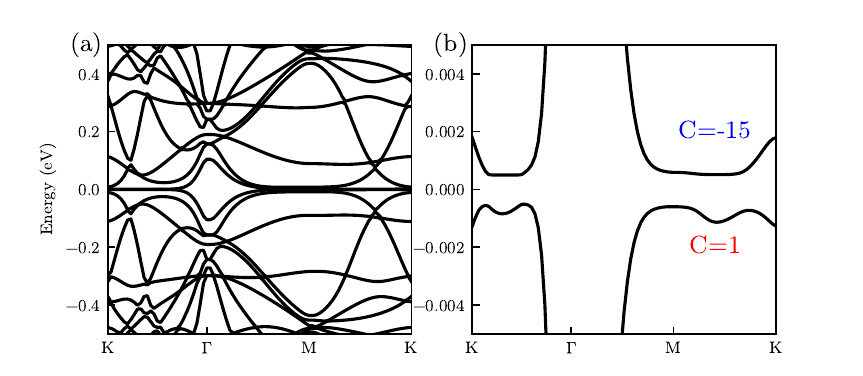}
    \caption{(a) Electronic band structure of 8+8-layer twisted rhombohedral graphene in the ABC-CBA configuration obtained using the continuum model. (b) Same band structure plotted in a narrow energy range showing the surface and interface bands with their corresponding Chern numbers.}
    \label{trg_band}
\end{figure}

In Fig. \ref{trg_band}, we present the band structure of 8+8-layer twisted rhombohedral graphene to illustrate the distinction between the surface and interface states. The two types of boundary modes have different origins. The surface states arise when the bulk is truncated by vacuum: the bulk–edge correspondence guarantees that the topological charge of the bulk phase must be discharged at the surface. 
In contrast, interface states appear at the domain boundaries between the regions characterized by different topological invariants: the mismatch enforces boundary-localized modes inside the bulk. In practice, the most reliable way to distinguish them is to compute their layer wavefunction localization.

\section{Band structure from tight-binding Green's function}\label{si_aa-ac}

\begin{figure}[htbp]
    \centering
    \includegraphics[width=8.6cm]{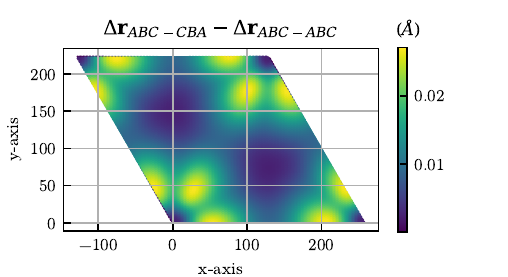}
    \caption{The displacement difference between ABC-CBA and ABC-ABC configurations, $\theta\approx0.5^\circ$.}
    \label{r_diff}
\end{figure}

% The transition from AA to AB is complete when reaching the chiral limit. 
The main difference between ABC-ABC and ABC-CBA configurations can be understood from the relaxation effect and the symmetry.
From Fig.~\ref{r_diff}, we observe a clear contrast in the relaxation patterns between the ABC–CBA and ABC–ABC configurations. Here $\Delta\mathbf{r}$ denotes the displacement between the unrelaxed and relaxed structures. The contrast clearly shows that the ABC–CBA stacking drives the middle TBG layers closer to the chiral limit, whereas ABC–ABC does not.

From a symmetry perspective, we can understand this behavior from the untwisted case by using an effective Green’s function,
\begin{equation}
    G_{ABC-CBA} = 
    \begin{pmatrix}
        2\Sigma(E) & h(\mathbf{k}) \\
        h^*(\mathbf{k}) &
    \end{pmatrix}^{-1}, 
    \qquad
        G_{ABC-ABC} = 
    \begin{pmatrix}
        \Sigma(E) & h(\mathbf{k}) \\
        h^*(\mathbf{k}) & \Sigma(E)
    \end{pmatrix}^{-1},
\end{equation}
where $\Sigma$ is the self-energy from the lead, and $h(\mathbf{k})=-t\phi(k_x,k_y)-\gamma_1e^{ik_zc}$. One can observe that the ABC-ABC configuration preserves the chiral symmetry with the energy shift. But for the ABC-CBA configuration, the self-energy is not equivalent for different sublattice sectors, which would provide a boundary mode without changing the bulk invariant. In this sense, the stacking fault would give us a flat band in rhombohedral graphite. When we introduce the twist stacking fault, this flat band would hybridize with the TBG flat bands, making the interface and surface flat bands flatten. 
% Let's introduce the chiral operator $\Gamma=\sigma_z$. The Hamiltonian would commute with the chiral operator $\Gamma$, $[G,\Gamma]=0$ as long as the $\Sigma_{L/R}=0$. Due to the twist effect, $\Sigma_L$ has the lattice mismatch with $\Sigma_R$. In this sense, $\mathbf{Tr}[G_{ABC-CBA},\Gamma]=2(\Sigma_L+\Sigma_R)$, which is smaller than $\mathbf{Tr}[G_{ABC-ABC},\Gamma]=2(\Sigma_L-\Sigma_R)$. 

\begin{figure}[htbp]
    \centering
    \includegraphics[width=14.6cm]{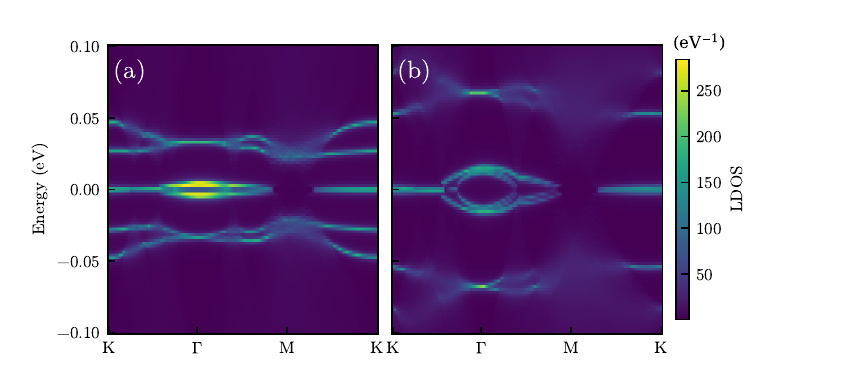}
    \caption{Momentum resolved DOS, solved from Green's function, twist angle is 2.13$^\circ$, (a) DOS for ABC-CBA configuration, (b) DOS for ABC-ABC configuration.}
    \label{aa-ac}
\end{figure}

% Each domain in ABC–CBA stacking is sublattice-polarized on opposite sites, so when they meet, the AA-type overlap between corresponding atoms is naturally weak. 
% OVY: I do not understand the first sentence as well as what is written below! This 
% XQ: Let's discuss if this is a good statement.
% However, the realistic situation must include the finite AA coupling. In the case of the ABC-CBA configuration, this sublattice polarization imposes a tighter constraint on the interface. 
We compared the momentum resolved DOS from two different configurations but the same twist angle, as shown in Fig. \ref{aa-ac}, the ABC-CBA configuration produced better flat bands.

\begin{figure}[htbp]
    \centering
    \includegraphics[width=8.6cm]{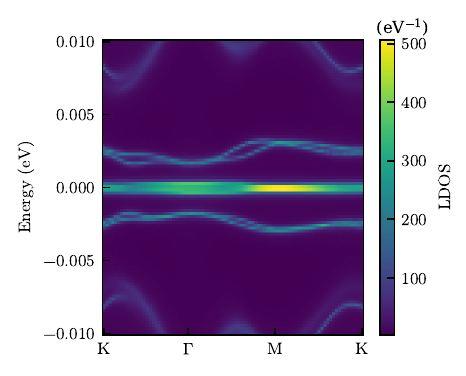}
    \caption{Momentum resolved DOS, ABC-CBA configuration, the twist angle is 1.08$^\circ$. }
    \label{ac30}
\end{figure}

We also compute the 'magic angle' situation, as we showed in Fig. \ref{ac30}, the 'chiral limit' case has the isolated flat bands with almost zero dispersion.

\end{document}